\documentclass[12pt,czech]{iopart}
\usepackage{iopams}
\usepackage[english]{babel}
\usepackage[T1]{fontenc}

\usepackage[utf8]{inputenc}
\usepackage{hyperref}
\usepackage{graphicx}
\usepackage{color}
\usepackage{amsthm}
\usepackage{amsfonts}
\usepackage{mathrsfs}
\usepackage{amssymb}
\usepackage{booktabs} 
\usepackage{multirow}
\usepackage{makecell} 
\usepackage{stackengine} 

\usepackage{etoolbox}
\preto\tabular{\shorthandoff{-}}

\newcommand{\itSigma}{\mathit{\Sigma}}
\newcommand{\calM}{\mathcal{M}}
\newcommand{\calN}{\mathcal{N}}
\newcommand{\ud}{\mathrm{d}}
\DeclareMathAlphabet{\mathpzc}{OT1}{pzc}{m}{it}
\newcommand{\calE}{\mathcal{E}}
\newcommand{\calZ}{\mathcal{Z}}

\newcommand{\ra}[1]{\renewcommand{\arraystretch}{#1}} 

\begin{document}

\title[Quasilocal horizons in inhomogeneous cosmological models]{Quasilocal horizons in inhomogeneous cosmological models}

\author{Eli\v{s}ka Pol\'{a}\v{s}kov\'{a} and Otakar Sv\'{\i}tek}
\address{Institute of Theoretical Physics, Faculty of Mathematics and Physics, Charles University, V~Hole\v{s}ovi\v{c}k\'{a}ch 2, 180~00 Prague 8, Czech Republic}
\ead{eli.polaskova@email.cz, ota@matfyz.cz}


\begin{abstract}
We investigate quasilocal horizons in inhomogeneous cosmological models, specifically concentrating on the notion of a trapping horizon defined by Hayward as a hypersurface foliated by marginally trapped surfaces. We calculate and analyse these quasilocally defined horizons in two dynamical spacetimes used as inhomogeneous cosmological models with perfect fluid source of non-zero pressure. In the spherically symmetric Lema\^{i}tre spacetime we discover that the horizons (future and past) are both null hypersurfaces provided that the Misner--Sharp mass is constant along the horizons. Under the same assumption we come to the conclusion that the matter on the horizons is of special character --- a perfect fluid with negative pressure. We also find out that they have locally the same geometry as the horizons in the Lema\^{i}tre--Tolman--Bondi spacetime. We then study the Szekeres--Szafron spacetime with no symmetries, particularly its subfamily with $\beta_{,z}\neq 0$, and we find conditions on the horizon existence in a general spacetime as well as in certain special cases.
\end{abstract}

\section*{Introduction}
Although standard cosmological models rely on the FLRW (Friedmann--Lema\^{\i}tre--Robertson--Walker) geometry, we know that our universe is highly inhomogeneous on scales smaller than the Hubble scale. The structure formation can then be accommodated either by perturbing the homogeneous model which leads to the interpretation of the homogeneous model as an averaged geometry (this is however far from being straightforward \cite{Ellis}) or by going beyond FLRW cosmologies towards inhomogeneous models. These issues are connected with the existence of dark energy as well. The possible substantial cosmological effects of treating inhomogeneity non-perturbatively were disputed recently (see, e.g. \cite{Green-Wald}) but opposing views exist that object to the general applicability of such results \cite{Buchert}. Nevertheless, no matter if the small-scale inhomogeneities affect the cosmological evolution substantially or not in the upcoming time of precision cosmology the exact treatment might be preferable over the perturbative approaches. Additional significance of inhomogeneous models is represented by their suitability for addressing local observational effects --- e.g. our position with respect to local filament/void structures and the ensuing modification of incoming radiation in our ''vicinity''.

The history of inhomogeneous models is long and we refer any interested reader to review publications on the topic \cite{Krasinski,Bolejko}. Here we are considering two specific families of geometries that admit cosmological fluid with pressure.

Lema\^{i}tre was the first to consider a spherically symmetric solution of the Einstein equations for a perfect fluid with non-zero pressure \cite{lemaitre}, and thus following \cite{alfedeel}, we refer to it as the Lema\^{i}tre spacetime. It is a generalization of the LTB (Lema\^{\i}tre--Tolman--Bondi) model for a perfect fluid with pressure. This solution was used, for example, to investigate whether the pressure gradients can prevent a shell-crossing singularity occuring in LTB \cite{bolejkolas}, or to calculate the effect of the inhomogeneity and pressure gradients on luminosity distance--redshift relations \cite{laskybol}.

The Szekeres--Szafron spacetime is a generalization of the Lema\^{i}tre spacetime. They both have the source in the form of a perfect fluid with pressure, but unlike the spherically symmetric Lema\^{i}tre solution, the Szekeres--Szafron model has no symmetries. This solution was first discovered by Szekeres \cite{szekeres} who considered only pressureless dust, and later generalized for a perfect fluid with pressure by Szafron \cite{szafron}. The Szekeres solution has been frequently used in cosmological applications: the effects on structure formation \cite{bolejko1,bolejko2} and the comparison with the standard perturbative treatment \cite{sussman}; the impact of the Szekeres--Swiss-cheese model on the propagation of light \cite{bolejko3} and the interpretation of the CMB observations \cite{bolejko4}; and the effect on a precision measurement of cosmological distances \cite{peel}. The averaging technique of Buchert was applied to the quasispherical Szekeres metric \cite{bolejko5}. The issue of avoiding shell-crossing singularity by properly setting up the solution using initial and final data was addressed using initial and final data \cite{walters} or solely initial data \cite{vrba}. Subsequently, realistic distributions of matter given by initial data for Szekeres spacetime that avoid shell-crossing singularities were investigated \cite{sussman2}

The standard approach to a black hole's boundary is represented by the event horizon. It is defined only in asymptotically well-behaved (i.e. flat, de Sitter or anti-de Sitter) spacetimes as the boundary of a region that cannot communicate with the future null infinity $\mathscr{I}^+$ \cite{hawking}. This definition works perfectly in stationary spacetimes, but in dynamical spacetimes it has some paradoxical properties due to its global nature. It can, for example, extend into flat regions of spacetime, and therefore predict the formation of a singularity (see pictures of the Vaidya spacetime in, e.g., \cite{senovilla, krishnan}). Moreover, this approach is sometimes inconvenient, namely in numerical relativity or initial value problem, where it is necessary to localize the horizon on a particular hypersurface without first evolving the whole spacetime.

A quasilocal approach to the boundary of a black hole addresses these problems caused by the global definition of an event horizon. In the 1970s, Hawking and Ellis were the first to introduce a horizon defined quasilocally, and they called it an \emph{apparent horizon} \cite{hawking}. Several more types of these horizons have been defined since then. In 1994, Hayward introduced a \emph{trapping horizon} \cite{hayward} and later, Ashtekar and Krishnan suggested the concept of an \emph{isolated horizon} (1999) and a \emph{dynamical horizon} (2002) \cite{ashtekar}. The apparent horizon, like the event horizon, had still considerable drawbacks \cite{hayward, jaramillo}, but Hayward's trapping horizon basically solved them. Ashtekar and Krishnan then defined an isolated horizon which describes a black hole in equilibrium with its neighbourhood, and a dynamical horizon which describes an evolving black hole, in order to obtain a quasilocal framework for deriving laws of black hole dynamics \cite{ashtekar}. In this paper, we will use the trapping horizon, however, we will also briefly mention its relation to the isolated and dynamical horizon.

The goal of this paper is to present an explicit calculation of quasilocal horizons (most authors call them apparent horizons with no reference to Hawking's definition) and, more importantly, to analyse their properties in the two aforementioned dynamical spacetimes, which are used as inhomogeneous cosmological models. 

Some work in this field has already been done. Quasilocal horizons in the Lema\^{i}tre spacetime have already been calculated by Alfedeel and Hellaby \cite{alfedeel} with the same results as we present in this paper. However on top of that, we thoroughly explore their geometrical properties.

Similarly in the Szekeres spacetime with dust, Hellaby and Krasiński \cite{hellkras} or Krasiński and Bolejko \cite{krasbol} have already calculated quasilocal horizons for the quasispherical geometry of the 2-surfaces of constant $t$ and $z$. Krasiński \cite{kras} also proved that there exist no apparent horizons in the case of quasiplanar and quasihyperbolic geometry of the 2-surfaces. In their calculations, all these authors assumed that the horizon was an orbit of a quasisymmetry, i.e. it was a surface \mbox{$\{t=\mathrm{const}, z=\mathrm{const}\}$} (which was sufficient to prove the non-existence of the horizon in the quasiplanar and quasihyperbolic case in \cite{kras}). However, here we aim to analyse the horizons in the generalized Szekeres--Szafron model with a perfect fluid and non-zero pressure, without any specific geometry of the 2-surfaces of constant $t$ and $z$ and without the assumption that the horizon section respects a quasisymmetry. Nevertheless, since the analysis of the most general case is very problematic, we also study a quasisymmetric horizon, thus generalizing the results of \cite{kras}.

As far as the notation is concerned, we consider a four-dimensional spacetime $\calM$ with the Lorentzian metric $g_{\mu\nu}$ with the signature $(-+++)$. The partial derivative is denoted by $_{,\alpha}\,$ and the covariant derivative by $_{;\alpha}\,$. The covariant derivative is compatible with the metric, i.e. $g_{\mu\nu;\alpha}=0$.

\section*{Double-null foliation, trapped surfaces and trapping horizon}
\begin{figure}[htb]
	\centering
	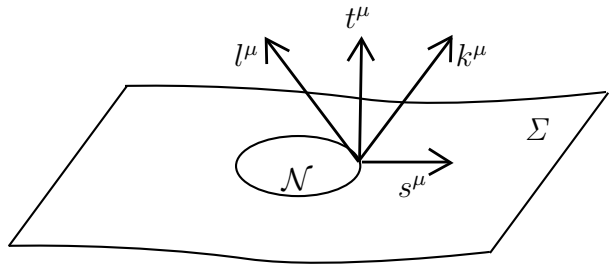
	\caption{\small{Double-null foliation of a spacetime with one temporal and two spatial dimensions.}}
	\label{fig:3+1}
\end{figure}

Suppose that a spacetime $\calM$ is foliated by spacelike hypersurfaces $\itSigma$ and let $t^{\mu}$ denote a timelike vector normal to $\itSigma$. Suppose there exists a closed (i.e. compact and without boundary) 2-surface $\calN$ on $\itSigma$, and let us denote a spacelike vector normal to $\calN$ lying on $\itSigma$ by $s^{\mu}$. These vectors satisfy
\begin{equation}
\label{eq:normy-ts}
t_{\mu}s^{\mu} = 0 \,,
\end{equation}
\begin{equation}
\label{eq:normy-tt}
t_{\mu}t^{\mu} = -1 \,,
\end{equation}
\begin{equation}
\label{eq:normy-ss}
s_{\mu}s^{\mu} = 1 \,,
\end{equation}
where the last two equations are our chosen normalizations. We define two future-directed null normal vectors to $\calN$ -- an outgoing one denoted by $k^{\mu}$, and an ingoing one denoted by $l^{\mu}$ -- as follows (see Figure \ref{fig:3+1})
\begin{eqnarray}
\label{eq:nullpomocits}
k^{\mu} & = t^{\mu}+s^{\mu} \,,\\
l^{\mu} & = t^{\mu}-s^{\mu} \,.
\end{eqnarray}
From this definition it follows that the normalization is $k_{\mu}l^{\mu}=-2$. The induced metric on $\calN$ then reads $h_{\mu\nu}=g_{\mu\nu}+\frac{1}{2}\left(k_{\mu}l_{\nu}+l_{\mu}k_{\nu}\right)$, and the expansion of a null normal $k^{\mu}$ is given by $\theta^{(k)}=h^{\mu\nu}k_{\mu;\nu}$. The foliation of a spacetime using two null vectors is referred to as the double-null foliation \cite{hayward}. We would like to point out that the double-null foliation can be equally constructed without defining a timelike and a spacelike vector first -- we can start directly with two future-directed null vectors that have the desired normalization and that are normal to a closed 2-surface.

Let us now define trapped surfaces and trapping horizons following mainly Hayward's paper \cite{hayward}, but \cite{jaramillo} offers a useful review, too. $\calN$ is said to be \emph{trapped} if $\theta^{(k)}\theta^{(l)}>0$, \emph{marginally trapped} if $\theta^{(k)}\theta^{(l)}=0$, and \emph{untrapped} if $\theta^{(k)}\theta^{(l)}<0$. A trapped surface is \emph{future} if $\theta^{(k)}<0$, $\theta^{(l)}<0$, and \emph{past} if $\theta^{(k)}>0$, $\theta^{(l)}>0$.

A \emph{trapping horizon} (TH) is defined as the closure of a hypersurface foliated by marginally trapped surfaces $\calN$ which satisfy
\begin{eqnarray}
\label{eq:TH-MTS}
\theta^{(k)} & = 0 \,,\\
\theta^{(l)} & \neq 0 \,,\\
\mathcal{L}_l\theta^{(k)} & \neq 0 \,,
\end{eqnarray}
where $\mathcal{L}_l$ denotes the Lie derivative along the vector $l^{\mu}$. We can choose the expansion of any null normal $k^{\mu}$ or $l^{\mu}$ to vanish, thus, for the purposes of this definition (and to be consistent with \cite{hayward}), we fixed $\theta^{(k)}=0$ on a horizon. Trapping horizons are classified according to the signs of $\theta^{(l)}$ and $\mathcal{L}_l\theta^{(k)}$. If on a horizon
\begin{eqnarray}
\label{eq:TH-futpast}
\theta^{(l)} & < 0 \,,\quad\textrm{ then TH is \emph{future}},\\
\theta^{(l)} & > 0 \,,\quad\textrm{ then TH is \emph{past}}.
\end{eqnarray}
However, in our notation we want to fix $k^{\mu}$ as an outgoing null normal and $l^{\mu}$ as an ingoing one, therefore a future trapping horizon will be given by the condition $\theta^{(k)}=0$, $\theta^{(l)}<0$, and a past trapping horizon will be given by $\theta^{(l)}=0$, $\theta^{(k)}>0$.
Furthermore, if on a (future) horizon
\begin{eqnarray}
\label{eq:TH-outin}
\mathcal{L}_l\theta^{(k)} & < 0 \,,\quad\textrm{ then TH is \emph{outer}},\\
\mathcal{L}_l\theta^{(k)} & > 0 \,,\quad\textrm{ then TH is \emph{inner}}.
\end{eqnarray}
The boundary of a black hole is thus described by a \emph{future outer trapping horizon}. One can include the case when $\mathcal{L}_l\theta^{(k)}=0$, and refer to such horizon as \emph{degenerate}. The same goes for a past horizon, only we evaluate the sign of $\mathcal{L}_k\theta^{(l)}$. 

However, it is not always possible to distinguish between an outgoing and an ingoing null normal vector due to lack of symmetries in a spacetime. This will be the case for the Szekeres--Szafron spacetime. In such cases, the quasilocal character of a trapping horizon proves useful. It is actually sufficient to know that one of the null normal congruences is non-expanding, the expansion of the other congruence is non-zero, and the Lie derivative of the first normal's expansion along the other normal vector is non-zero as well, and there is no need to specify which normal is which.

Comparing the definition of a trapping horizon with the properties of isolated and dynamical horizons, which are thoroughly discussed in \cite{ashtekar}, we can see that a null trapping horizon (future or past) is the same thing as a non-expanding horizon, and a spacelike future trapping horizon is the same thing as a dynamical horizon.

\section*{Lema\^{i}tre spacetime}
This spherically symmetric solution of the Einstein equations has a perfect fluid with non-zero pressure as a source and was first considered by Lema\^{i}tre \cite{lemaitre}. In the subsequent overview, we follow \cite{plebkras}. The metric in the so-called comoving-synchronous coordinates takes the form
\begin{equation}
\label{eq:lemaitre}
g_{\mu\nu}\ud x^{\mu}\ud x^{\nu} = -\rme^{C(t,r)}\ud t^2+\rme^{A(t,r)}\ud r^2+R^2(t,r)\left(\ud\vartheta^2+\sin^2\vartheta\,\ud\varphi^2\right) \,,
\end{equation}
where $C(t,r)$ and $A(t,r)$ are functions to be specified, and $R(t,r)$ is the areal radius of a sphere $\{t=\mathrm{const}, r=\mathrm{const}\}$. The 4-velocity of the fluid is $u^{\mu}=\rme^{-\frac{C}{2}}\delta^{\mu}_0$, and its norm is $u_{\mu}u^{\mu}=-1$, thus it plays the role of the timelike vector $t^{\mu}$ defined in the previous section by \eref{eq:normy-tt}. The Einstein equations with the cosmological constant taken into account read
\begin{equation}
\kappa p = -\frac{2M_{,t}}{R^2R_{,t}} \,, \label{eq:lem-EEq-tlak}
\end{equation}
\begin{equation}
\kappa\varepsilon = \frac{2M_{,r}}{R^2R_{,r}} \,, \label{eq:lem-EEq-en}
\end{equation}
where
\begin{equation}
\label{eq:lem-M}
2M(t,r) = R+\rme^{-C}RR_{,t}{}^2-\rme^{-A}RR_{,r}{}^2-\frac{1}{3}\Lambda R^3 \,,
\end{equation}
$p$ denotes pressure and $\varepsilon$ energy density. Generally, the pressure $p(t,r)$ and the energy density $\varepsilon(t,r)$ are functions of both coordinates $t$ and $r$. The function $M(t,r)$ is referred to as the Misner--Sharp mass, and it has all the attributes of physical mass. The limit case of zero pressure leads to the well-known Lema\^{i}tre--Tolman--Bondi (LTB) solution, which is the simplest inhomogeneous cosmological model. If we restrict ourselves to the case when the pressure depends only on time, then it follows from the conservation laws $T^{\mu\nu}{}_{;\nu}=0$ that $C_{,r}=0$, and we can rescale $t$ so that $g_{tt}=-1$, and thus the metric obtains a form similar to LTB without pressure. This case of LTB spacetime with pressure was thoroughly discussed in a recent paper by Lynden-Bell and Bičák \cite{bicak}.

\subsection*{Future and past horizon}
Because of the spherical symmetry of the Lema\^{i}tre spacetime, we consider the horizon to be spherically symmetric as well, thus independent of the angular coordinates $\vartheta$ and $\varphi$, and therefore the complete horizon hypersurface is described by $\rho(t,r)=0$. However, we consider only a spatial slice of the horizon, i.e. $t=t_0$. The equation describing the 2-surface of the horizon then takes the form
\begin{equation}
\label{eq:lem-ro}
\rho(t_0, r) = 0 \,.
\end{equation}
A 1-form normal to the horizon section is obtained by using gradient of \eref{eq:lem-ro}, and transformed into the null normal 1-form $k_{\mu}$ by linear combination with $\ud t$. From $k_{\mu}k^{\mu}=0$ we obtain the following null normal 1-form and the corresponding vector
\begin{eqnarray}
\label{eq:lem-kkvect}
k_{\mu}\ud x^{\mu} & = \rho_{,r}\left(-\rme^{\frac{C-A}{2}}\,\ud t + \ud r\right) \,,\\[0.1cm]
k^{\mu}\frac{\partial}{\partial x^{\mu}} & = \rho_{,r}\left(\rme^{-\frac{C+A}{2}}\,\frac{\partial}{\partial t} + \rme^{-A}\,\frac{\partial}{\partial r}\right) \,.
\end{eqnarray}
The vector is future-directed provided that $\rho_{,r}>0$. Otherwise, if $\rho_{,r}=0$, the null normal would be $k^{\mu}\equiv 0$, and there would be no horizon. Such situation is not interesting for us, thus $\rho_{,r}$ must always have the same sign which we choose positive. Then the vector $k^{\mu}$ is a future-directed \emph{outgoing} null normal. Changing the sign of the radial component of $k_{\mu}$ and using the normalization condition, a future-directed \emph{ingoing} null normal 1-form and the corresponding vector are obtained 
\begin{eqnarray}
\label{eq:lem-llvect}
l_{\mu}\ud x^{\mu} & = \frac{1}{\rho_{,r}}\left(-\rme^{\frac{C+A}{2}}\,\ud t - \rme^{A}\ud r\right) \,,\\[0.1cm]
l^{\mu}\frac{\partial}{\partial x^{\mu}} & = \frac{1}{\rho_{,r}}\left(\rme^{-\frac{C-A}{2}}\,\frac{\partial}{\partial t} - \frac{\partial}{\partial r}\right) \,.
\end{eqnarray}
The induced metric is then
\begin{equation}
\label{eq:lem-h}
h_{\mu\nu}\ud x^{\mu}\ud x^{\nu} = R^2\,\ud\vartheta^2+R^2\sin^2\vartheta\,\ud\varphi^2 \,,
\end{equation}
and thus expansions of the null normals take the form
\begin{eqnarray}
\label{eq:lem-expanze}
\theta^{(k)} & = \frac{2\rho_{,r}}{R}\left(\rme^{-\frac{C+A}{2}}R_{,t}+\rme^{-A}R_{,r}\right) \,,\\[0.1cm]
\theta^{(l)} & = \frac{2}{R\,\rho_{,r}}\left(\rme^{-\frac{C-A}{2}}R_{,t}-R_{,r}\right) \,.
\end{eqnarray}

The condition $\theta^{(k)}=0$ for a future horizon gives the following equation
\begin{equation}
\label{eq:lem-rcehor}
R_{,t}+\rme^{\frac{C-A}{2}}R_{,r} = 0 \,.
\end{equation}
Using this equation, we get $\theta^{(l)}=-\frac{4R_{,r}}{R\,\rho_{,r}}<0$, because $R>0$ (positive areal radius), $R_{,r}>0$ (positive sign is a natural choice to prevent a shell-crossing singularity $R_{,r}=0$ -- the radius of a sphere \mbox{$\{t=\mathrm{const}, r=\mathrm{const}\}$} grows with the increasing radial coordinate $r$), and $\rho_{,r}>0$. Thus the horizon given by \eref{eq:lem-rcehor} is future indeed. Moreover, this equation has a non-trivial solution for $R(t,r)$ if and only if $R_{,t}<0$ -- in other words, the future horizon exists only in a collapsing phase of the universe\footnote{We adopt the terminology of a "collapsing" and an "expanding" phase of the universe as referring only to the sign of $R_{,t}$, following e.g. \cite{hellkras}.}.

From $\theta^{(l)}=0$ on a past horizon it follows that
\begin{equation}
\label{eq:lem-minhor}
R_{,t} - \rme^{\frac{C-A}{2}}R_{,r} = 0 \,.
\end{equation}
Using this equation, we get $\theta^{(k)}=\frac{4\,\rme^{-A}R_{,r}\rho_{,r}}{R}>0$, which confirms that the horizon given by \eref{eq:lem-minhor} is past. In this case, the past horizon exists only in an expanding phase of the universe which is characterized by the condition $R_{,t}>0$.

To verify these results, one can apply the limit of zero pressure as in \cite{plebkras} to equations \eref{eq:lem-rcehor} and \eref{eq:lem-minhor}, and obtain the equations of future and past horizon in LTB spacetime.

\subsection*{Adapted coordinates}
Since $C(t,r)$, $A(t,r)$ and $R(t,r)$ are general functions and equations \eref{eq:lem-rcehor} and \eref{eq:lem-minhor} are algebraic (we fixed the time coordinate $t$, and from \eref{eq:lem-ro} we are able to obtain a constant solution for $r$), we introduce new coordinates $\xi(t,r)$ and $\eta(t,r)$ adapted to the equations of future and past horizon
\begin{eqnarray}
\label{eq:lem-xieta}
\xi & = t+F(t,r)\,r \,,\\
\eta & = t-F(t,r)\,r \,,
\end{eqnarray}
with $F(t,r)$ being an arbitrary function. Both equations of future and past horizon adopt the form $R_{,\xi}=0$, provided that $F(t,r)$ satisfies
\begin{equation}
\label{eq:lem-F}
1 = F_{,t}\,r \pm \rme^{\frac{C-A}{2}}\left(F_{,r}\,r+F\right) \,,
\end{equation}
where the sign ''$+$'' applies in the case of future horizon, and ''$-$'' applies in the case of past horizon. The advantage of these coordinates is that both horizons develop in the direction of the coordinate $\xi$, as we will show below. Rewriting equation \eref{eq:lem-M} into these new coordinates on the horizon, we obtain
\begin{equation}
\label{eq:lem-M-xieta}
2M(\xi,\eta) = R - \frac{1}{3}\Lambda R^3 \,.
\end{equation}
Using \eref{eq:lem-M}, we obtain from the Einstein equations \eref{eq:lem-EEq-tlak} and \eref{eq:lem-EEq-en} in new coordinates the equivalence
\begin{equation}
\label{eq:lem-stavovarce}
\varepsilon = -p \,\quad\Leftrightarrow\,\quad M_{,\xi} = 0 \,,
\end{equation}
which holds only on the horizons, where $R_{,\xi}=0$. The first equation is the equation of state corresponding to the cosmological constant acting as a source. This means that the matter present on the horizons is rather special, because it behaves locally as a perfect fluid with negative pressure whose magnitude is equal to the energy density. However, this is true locally if and only if the Misner–Sharp mass is constant along the given portion of horizon.

Note however that the above condition is only local since it was derived using a quasilocal horizon definition. If it is satisfied, we have an isolated horizon defined specifically to allow for non-trivial matter content in the vicinity of the horizon while it should not cross it (otherwise it becomes dynamic and spatial). If we would assume a fluid that is present everywhere in the spacetime and satisfies fixed equation of state, the above condition would imply Schwarzschild--de Sitter solution. However, allowing for models leading to the collapse of fluid, the horizon portion to the future of the point where all the fluid passed through it would exactly satisfy the condition \eref{eq:lem-stavovarce}. This point is analyzed in the case of a specific generalisation of LTB spacetime with pressure \cite{laskylun} when the apparent horizons are discussed. In these cases we would have trivially $\varepsilon=-p=0$, but considering an additional ''cosmological constant'' fluid component we can obtain non-zero values. This would also bring the possibility to apply our conditions on the cosmological horizon provided the standard fluid component is restricted to be inside from a certain moment in time.

\subsection*{Properties of the horizons}
In this section, we determine under what conditions the horizons are outer and inner, we prove that they are both null (provided that $M_{,\xi}=0$) and have locally the same geometry as the horizons in LTB spacetime, and we also provide evidence for the matter on the horizons being of special character when $M_{,\xi}=0$.

In order to determine whether the horizons are outer or inner, a ''cross-focusing'' equation will be used (it describes the cross-focusing between the two null congruences, see Hayward \cite{hayward}). For non-twisting null congruences, which is our case because both congruences are hypersurface orthogonal, and for the future horizon it takes the form
\begin{equation}
\label{eq:crossraychaud}
\mathcal{L}_l\theta^{(k)} + \theta^{(k)}\theta^{(l)} + \frac{1}{4}\mathcal{R} = \kappa T_{\mu\nu}k^{\mu}l^{\nu} \,,
\end{equation}
with $\mathcal{R}$ being the Ricci scalar of the horizon section. For the past horizon we just interchange the outgoing and the ingoing vector $k^{\mu}\longleftrightarrow l^{\mu}$, and we get the same equation up to the first term which is $\mathcal{L}_k\theta^{(l)}$. If the dominant energy condition \cite{wald} holds, then the right-hand side is always non-negative. In our case, using the stress-energy tensor for a perfect fluid in the form $T_{\mu\nu}=(\varepsilon+p)u_{\mu}u_{\nu}+pg_{\mu\nu}$, the right-hand side can be expressed as
\begin{equation}
\label{eq:lem-Tkl}
\kappa T_{\mu\nu}k^{\mu}l^{\nu} = \kappa\left[(\varepsilon+p)u_{\mu}k^{\mu}u_{\nu}l^{\nu} + pg_{\mu\nu}k^{\mu}l^{\nu}\right] = \kappa(\varepsilon-p) \geq 0 \,.
\end{equation}
The inequality holds provided that $\varepsilon\geq 0$ and $\varepsilon\geq|p|$, which is equivalent to the dominant energy condition satisfied. The Ricci scalar of the horizons is $\mathcal{R}=\frac{2}{R^2(t,r)}>0$. All put together, the horizons are
\begin{equation}
\label{eq:lem-out}
\textrm{outer}\,\quad\Leftrightarrow\,\quad \kappa\varepsilon < \frac{1}{4R^2}\,,
\end{equation}
\begin{equation}
\label{eq:lem-in}
\textrm{inner}\,\quad\Leftrightarrow\,\quad \kappa\varepsilon > \frac{1}{4R^2}\,.
\end{equation}
There exists only one future and one past horizon since \eref{eq:lem-M-xieta} has only one real solution, therefore both horizons are either outer or inner.

The character of the future and the past horizon follows from the sign of $\frac{\mathcal{L}_k\theta^{(k)}}{\mathcal{L}_l\theta^{(k)}}$ and $\frac{\mathcal{L}_l\theta^{(l)}}{\mathcal{L}_k\theta^{(l)}}$, respectively \cite{hayward}. The Raychaudhuri equation for non-twisting and non-shearing (which is our case so that the spherical symmetry is preserved) affinely parametrized null congruences reads
\begin{eqnarray}
\label{eq:raychaud2}
\mathcal{L}_k\theta^{(k)} & = -\kappa T_{\mu\nu}k^{\mu}k^{\nu} \,,\\
\mathcal{L}_l\theta^{(l)} & = -\kappa T_{\mu\nu}l^{\mu}l^{\nu} \,,
\end{eqnarray}
on the future and on the past horizon, respectively. Consequently, one arrives at the following expression
\begin{equation}
\mathcal{L}_k\theta^{(k)} = \mathcal{L}_l\theta^{(l)} = -\kappa(\varepsilon+p) = 0 \,\quad\Leftrightarrow\,\quad M_{,\xi} = 0 \,,
\end{equation}
due to \eref{eq:lem-stavovarce}. Thus both horizons are null if and only if the Misner--Sharp mass is constant along the horizons. This means that in this special case the direction of the evolution of the future and the past horizon is given exactly by the vector $k^{\mu}$ and $l^{\mu}$, respectively. Using the adapted coordinates introduced above and the corresponding condition on the function $F(t,r)$ \eref{eq:lem-F}, one finds out that both horizons develop in the direction of the coordinate $\xi$ as both vectors $k^{\mu}\frac{\partial}{\partial x^{\mu}}$ and $l^{\mu}\frac{\partial}{\partial x^{\mu}}$ are proportional to $\frac{\partial}{\partial\xi}$. This also means that they are both non-expanding horizons according to \cite{ashtekar} (we refer the reader to this reference for more details on the properties of non-expanding horizons, some of which we also use in the next paragraph).

We will now provide evidence for the equation of state on the horizons \eref{eq:lem-stavovarce} being mathematically and physically correct (when $M_{,\xi}=0$). Since the condition $\frac{1}{2}R_{\mu\nu}k^{\mu}k^{\nu}=0$ is satisfied on a non-expanding horizon \cite{ashtekar}, we can rewrite it using the Einstein equations and the stress-energy tensor for a perfect fluid into the form
\begin{equation}
\label{eq:lem-Riccicomp}
0 = \frac{1}{2}R_{\mu\nu}k^{\mu}k^{\nu} = \frac{1}{2}\kappa(\varepsilon+p) \,.
\end{equation}
This is true if and only if \eref{eq:lem-stavovarce} holds, therefore there is no ordinary matter on the horizons (there is a perfect fluid with negative pressure instead). Physically, this can be explained in the following way. 
Since a non-expanding horizon is in equilibrium with its neighbourhood, no matter with non-zero energy density can cross it, and thus the condition $M_{,\xi}=0$ is satisfied. Therefore, particles could in theory follow only quasicircular orbits around the horizon. However, because of the above result, there cannot be ordinary matter in the surroundings of the horizons, which is analogous to the non-existence of stable circular orbits for (dust) particles in the vicinity of the horizon in the Schwarzschild--de Sitter spacetime. And indeed, substituting $p=0$ for dust particles into equation \eref{eq:lem-Riccicomp}, it follows that they have necessarily zero energy density $\varepsilon=0$.

Finally, comparing equation \eref{eq:lem-M-xieta} with the equation of the horizon in LTB spacetime (see e.g. \cite{hellaby}), one finds out that they have identical form. Let us now compare intrinsic and extrinsic geometry of the horizons in these spacetimes. First of all, both spacetimes are spherically symmetric, and calculating the induced metric in LTB spacetime, one concludes that indeed they have the same form \eref{eq:lem-h}. Moreover, the function $R(t,r)$ represents in both cases the areal radius of a sphere $\{t=\mathrm{konst},r=\mathrm{konst}\}$, therefore the intrinsic geometry is identical. As for the extrinsic geometry, one can compare extrinsic curvature tensors in a null direction $k^{\mu}$, given by $\theta_{\mu\nu}^{(k)}=h_{\mu}{}^{\rho}h_{\nu}{}^{\sigma}k_{(\rho;\sigma)}$. Calculating them in both null directions normal to the horizon, we arrive at the following expressions on the future horizon in the Lema\^{i}tre spacetime
\begin{eqnarray}
\theta^{(k)}_{\mu\nu}\ud x^{\mu}\ud x^{\nu} & = 0 \,,\\
\theta^{(l)}_{\mu\nu}\ud x^{\mu}\ud x^{\nu} & = -\frac{2RR_{,r}}{\rho_{,r}}
\left(\ud\vartheta^2+\sin^2\vartheta\,\ud\varphi^2\right) \,.
\end{eqnarray}
The same formulae are obtained on the future horizon in LTB spacetime. As for the past horizon in the Lema\^{i}tre spacetime, the extrinsic curvature tensors are
\begin{eqnarray}
\theta^{(k)}_{\mu\nu}\ud x^{\mu}\ud x^{\nu} & = 2\,\rme^{-A}RR_{,r}\,\rho_{,r} \left(\ud\vartheta^2+\sin^2\vartheta\,\ud\varphi^2\right) \,,\\
\theta^{(l)}_{\mu\nu}\ud x^{\mu}\ud x^{\nu} & = 0 \,,
\end{eqnarray}
and after applying the limit transition, we obtain the identical expressions for the past horizon in LTB spacetime. Thus the horizons in the Lema\^{i}tre spacetime have locally the same geometry as the horizons in LTB spacetime. This confirms the result in \cite{alfedeel}, where the horizon was calculated as the locus where an observer's past null cone reaches its maximum areal radius, i.e. $R_{,r}=0$ along an incoming radial null geodesic.

\section*{Szekeres--Szafron spacetime}
This solution of the Einstein equations has also a perfect fluid with non-zero pressure as a source, but unlike the Lema\^{i}tre spacetime, it has no symmetries. In this overview, we also follow \cite{plebkras}. The metric in the comoving coordinates takes the form
\begin{equation}
\label{eq:ss}
g_{\mu\nu}\ud x^{\mu}\ud x^{\nu} = -\ud t^2 + \rme^{2\beta}\left(\ud x^2 + \ud y^2\right) + \rme^{2\alpha}\ud z^2 \,,
\end{equation}
with $\alpha$ and $\beta$ being functions of all the coordinates. The comoving coordinates imply $u^{\mu}=\delta^{\mu}_{0}$, therefore $\dot{u}^{\mu}=0$, and the pressure $p(t)$ depends only on time. Also, $u_{\mu}u^{\mu}=-1$, thus $u^{\mu}$ again plays the role of the timelike vector $t^{\mu}$ \eref{eq:normy-tt} from before. There are two subfamilies of these solutions, depending on whether $\beta_{,z}=0$ or $\beta_{,z}\neq 0$. In the following, we are considering only the case $\beta_{,z}\neq 0$, which includes the Lema\^{i}tre spacetime as a spherically symmetric limit. For more details about both subfamilies see e.g. \cite{plebkras}.

The Einstein equations lead to the following expressions for the metric functions for this subfamily 
\begin{eqnarray}
	\rme^{\beta} & = \Phi(t,z)\rme^{\nu(x,y,z)} \,, \label{eq:ss-beta} \\
	\rme^{\alpha} & = h(z)\rme^{-\nu(x,y,z)}\left(\rme^{\beta}\right)_{,z} \,, \label{eq:ss-alpha}
\end{eqnarray}
and yield the following restrictions on the functions $\Phi(t,z)$ and $\rme^{\nu(x,y,z)}$
\begin{equation}
\label{eq:ss-phi}
2\,\frac{\Phi_{,tt}}{\Phi} + \frac{\Phi_{,t}{}^2}{\Phi^2} + \kappa p(t) - \Lambda + \frac{k(z)}{\Phi^2} = 0 \,,
\end{equation}
\begin{equation}
\label{eq:ss-nu}
\rme^{-\nu(x,y,z)} = A(z)(x^2+y^2) + 2B_1(z)\,x + 2B_2(z)\,y + C(z) \,,
\end{equation}
where the functions $A(z)$, $B_1(z)$, $B_2(z)$, $C(z)$, $h(z)$ and $k(z)$ satisfy
\begin{equation}
\label{eq:ss-ABChk}
AC-B_1{}^2-B_2{}^2 = \frac{1}{4}\left(\frac{1}{h^2(z)}+k(z)\right) \,,
\end{equation}
otherwise they are arbitrary. The sign of $g(z)=AC-B_1{}^2-B_2{}^2$ determines the geometry of the surfaces {$\{t=\textrm{const},z=\textrm{const}\}$}. However, the character of these surfaces can change for different $z$ within a single $t=\textrm{const}$ hypersurface.
Equation \eref{eq:ss-phi} can be formally integrated to obtain 
\begin{equation}
\label{eq:ss-phiint}
\Phi_{,t}{}^2 = \frac{1}{3}\Lambda\Phi^2 - k(z) + \frac{2M(z)}{\Phi} - \frac{\kappa}{3\Phi}\int p(t)\left(\Phi^3\right)_{,t} \ud t \,.
\end{equation}

However, we will use a different parametrization of the metric functions following \cite{kras}. Let us define new functions
\begin{eqnarray}
\label{eq:ss-reparam}
(A,B_1,B_2) = \frac{\sqrt{|g|}}{2S}(1,-P,-Q) \,, \qquad \epsilon = \frac{g(z)}{|g(z)|} \,, \\
k(z) = -|g(z)|\cdot 2E(z) \,, \qquad M(z) = |g(z)|^{3/2}\widetilde{M}(z) \,, \\
\Phi(t,z) = R(t,z)\sqrt{|g(z)|} \,.
\end{eqnarray}
Then the metric \eref{eq:ss} becomes
\begin{equation}
\label{eq:ssreparam}
g_{\mu\nu}\ud x^{\mu}\ud x^{\nu} = -\ud t^2 + \frac{R^2}{\calE^2}\left(\ud x^2 + \ud y^2\right) + \frac{\left(R_{,z}-R\,\displaystyle{\frac{\calE_{,z}}{\calE}}\right)^2}{\epsilon+2E}\,\ud z^2 \,,
\end{equation}
where
\begin{equation}
\label{eq:ss-calE}
\frac{\rme^{-\nu(x,y,z)}}{\sqrt{|g(z)|}} = \calE(x,y,z) = \frac{S}{2}\left[\left(\frac{x-P}{S}\right)^2+\left(\frac{y-Q}{S}\right)^2+\epsilon\right] \,,
\end{equation}
and equation \eref{eq:ss-phi} becomes
\begin{equation}
\label{eq:ss-R}
2\,\frac{R_{,tt}}{R} + \frac{R_{,t}{}^2}{R^2} + \kappa p(t) - \Lambda - \frac{2E(z)}{R^2} = 0 \,.
\end{equation}
The parameter $\epsilon$ (not to confuse with the energy density $\varepsilon$) can have values $+1, 0$ or $-1$, which correspond to the surfaces $\{t=\textrm{const},z=\textrm{const}\}$ being spherical, planar or hyperbolic, respectively. Following \cite{hellkras}, we assume that the function $R(t,z)\geq 0$, because it represents the areal radius. 
If $R_{,z}=R\frac{\calE_{,z}}{\calE}$, we get a shell-crossing singularity $\varepsilon\rightarrow\infty$. Therefore, we need to distinguish two possibilities to avoid it -- either $R_{,z}-R\frac{\calE_{,z}}{\calE}<0$ or $R_{,z}-R\frac{\calE_{,z}}{\calE}>0$. Finally, in order to preserve the Lorentzian signature of the metric, we assume that $\epsilon+2E>0$.

\subsection*{Future and past horizon}
\begin{figure}[htb]
	\centering
	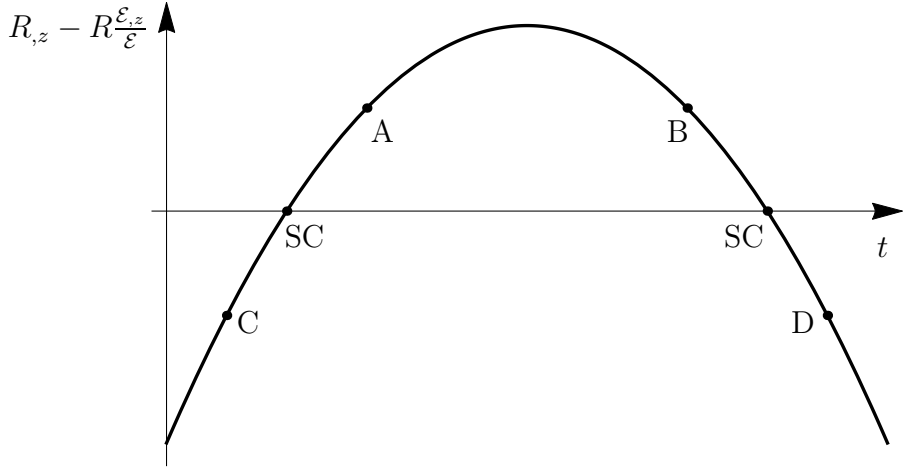
	\caption{\small{A possible time evolution of the neighbouring matter shells distance (an artistic illustration). The distance is locally given by $\pm\sqrt{g_{zz}}=\frac{R_{,z}-R\,\frac{\calE_{,z}}{\calE}}{\sqrt{\epsilon+2E}}$ with the denominator always positive, so it does not affect the sign (by a negative sign we only mean that the shells have been reordered). Two points denoted by SC represent shell-crossing singularities with $R_{,z}=R\frac{\calE_{,z}}{\calE}$ and diverging energy density. This occurs when the neighbouring matter shells collide. At A, the distance between the neighbouring matter shells is positive as $R_{,z}-R\frac{\calE_{,z}}{\calE}\Big|_A>0$, and the function is further increasing, i.e. $\left(R_{,z}-R\frac{\calE_{,z}}{\calE}\right)_{\!\!,t}\bigg|_A>0$, therefore the shells are moving away from each other. On the other hand, the function is decreasing at B, i.e. $\left(R_{,z}-R\frac{\calE_{,z}}{\calE}\right)_{\!\!,t}\bigg|_B<0$, and so the matter shells are moving towards each other. At C and D, the distance between the neighbouring shells is negative since $R_{,z}-R\frac{\calE_{,z}}{\calE}\Big|_{C,D}<0$, and so it is the other way round: At C, the matter shells are moving towards each other, whereas at D, they are moving away from each other.}}
	\label{fig:shellcross}
\end{figure}

Since the spacetime has no symmetries, we take the equation describing a spatial slice ($t=t_{0}$) of the horizon in the most general form
\begin{equation}
\label{eq:ss-ro}
\rho(t_0,x,y,z) = \textrm{const} > 0 \,.
\end{equation}
Let us first consider the case $R_{,z}-R\frac{\calE_{,z}}{\calE}<0$. Calculating null normal 1-forms and the corresponding vectors in the same way as for the Lema\^{i}tre spacetime, one gets
\begin{eqnarray}
\label{eq:ss-reparam-kkvect}
k_{\mu}\ud x^{\mu}  = & -\frac{\sqrt{\left(R_{,z}\calE-R\calE_{,z}\right)^2\left(\rho_{,x}{}^2+\rho_{,y}{}^2\right)+R^2(\epsilon+2E)\rho_{,z}{}^2}}
{R\left(R\frac{\calE_{,z}}{\calE}-R_{,z}\right)}\,\ud t\\
& + \rho_{,x}\,\ud x + \rho_{,y}\,\ud y + \rho_{,z}\,\ud z \,, \nonumber \\[0.5cm]
k^{\mu}\frac{\partial}{\partial x^{\mu}} = & \frac{\sqrt{\left(R_{,z}\calE-R\calE_{,z}\right)^2\left(\rho_{,x}{}^2+\rho_{,y}{}^2\right)+R^2(\epsilon+2E)\rho_{,z}{}^2}}
{R\left(R\frac{\calE_{,z}}{\calE}-R_{,z}\right)}\,\frac{\partial}{\partial t} \\
& + \frac{\calE^2}{R^2}\left(\rho_{,x}\,\frac{\partial}{\partial x} + \rho_{,y}\,\frac{\partial}{\partial y}\right) + \frac{\epsilon+2E}{\left(R\frac{\calE_{,z}}{\calE}-R_{,z}\right)^2}\,\rho_{,z}
\,\frac{\partial}{\partial z} \,, \nonumber
\end{eqnarray}
which is a future-directed null normal, because $R_{,z}-R\frac{\calE_{,z}}{\calE}<0$. All the derivatives $\rho_{,x}$, $\rho_{,y}$ and $\rho_{,z}$ being non-zero is the most general case reflecting the fact that the spacetime and the horizon have no symmetries; otherwise the horizon would be symmetric with respect to the corresponding coordinate(s). Also, due to this absence of symmetries, there is no way to distinguish between an outgoing and an ingoing direction. Therefore, we can assume that $\rho_{,x}>0$, $\rho_{,y}>0$ and $\rho_{,z}>0$ without loss of generality. However, we make this assumption at a generic point in the spacetime, therefore there can exist points with some of the derivatives equal to zero. The other future-directed null normal 1-form and the corresponding vector are then
\begin{eqnarray}
\label{eq:ss-reparam-llvect}
l_{\mu}\ud x^{\mu} = -\frac{R\left(R\frac{\calE_{,z}}{\calE}-R_{,z}\right)}{\sqrt{\left(R_{,z}\calE-R\calE_{,z}\right)^2\left(\rho_{,x}{}^2+\rho_{,y}{}^2\right)+R^2(\epsilon+2E)\rho_{,z}{}^2}}\,\ud t& \\
-\frac{R^2\left(R\frac{\calE_{,z}}{\calE}-R_{,z}\right)^2}{\left(R_{,z}\calE-R\calE_{,z}\right)^2\left(\rho_{,x}{}^2+\rho_{,y}{}^2\right)+R^2(\epsilon+2E)\rho_{,z}{}^2} \left(\rho_{,x}\,\ud x + \rho_{,y}\,\ud y + \rho_{,z}\,\ud z\right)\,,& \nonumber\\[0.7cm]
l^{\mu}\frac{\partial}{\partial x^{\mu}} = \frac{R\left(R\frac{\calE_{,z}}{\calE}-R_{,z}\right)}{\sqrt{\left(R_{,z}\calE-R\calE_{,z}\right)^2\left(\rho_{,x}{}^2+\rho_{,y}{}^2\right)+R^2(\epsilon+2E)\rho_{,z}{}^2}} \,\frac{\partial}{\partial t}& \\[0.1cm]
-\frac{\left(R_{,z}\calE-R\calE_{,z}\right)^2}{\left(R_{,z}\calE-R\calE_{,z}\right)^2\left(\rho_{,x}{}^2+\rho_{,y}{}^2\right)+R^2(\epsilon+2E)\rho_{,z}{}^2}
\left(\rho_{,x}\,\frac{\partial}{\partial x} + \rho_{,y}\,\frac{\partial}{\partial y}\right) &\nonumber\\[0.2cm]
-\frac{R^2(\epsilon+2E)}{\left(R_{,z}\calE-R\calE_{,z}\right)^2\left(\rho_{,x}{}^2+\rho_{,y}{}^2\right)+R^2(\epsilon+2E)\rho_{,z}{}^2}
\,\rho_{,z}\,\frac{\partial}{\partial z} \,.&\nonumber
\end{eqnarray}
Expansions of the null normals can be expressed as
\begin{equation}
\label{eq:ss-reparam-expanzek}
\theta^{(k)} = \Delta\rho(t_0,x,y,z)+\calZ(t_{0},x,y,z) \,,
\end{equation}
\begin{equation}
\label{eq:ss-reparam-expanzel}
\theta^{(l)} = N\left\{-\Delta\rho(t_0,x,y,z)+\calZ(t_{0},x,y,z)\right\} \,,
\end{equation}
where the Laplace operator is defined with the induced metric, i.e. $\Delta\rho(t_0,x,y,z)=h^{\mu\nu}\rho_{;\mu\nu}$. We denoted by $\calZ(t_{0},x,y,z)$ the expression
\begin{equation}
\label{eq:ss-reparam-Z}
\calZ = \frac{-R\calE^2\!\!\left(R_{,z}-R\frac{\calE_{,z}}{\calE}\right)\!\!\left(R_{,z}-R\frac{\calE_{,z}}{\calE}\right)_{\!\!,t}\!\left(\rho_{,x}{}^2+\rho_{,y}{}^2\right)\!+R_{,t}\,\Xi}{R^2\left(R_{,z}-R\frac{\calE_{,z}}{\calE}\right)\sqrt{\left(R_{,z}\calE-R\calE_{,z}\right)^2\left(\rho_{,x}{}^2+\rho_{,y}{}^2\right)+R^2(\epsilon+2E)\rho_{,z}{}^2}}
\end{equation}
and introduced $\Xi=\left[\left(R_{,z}\calE-R\calE_{,z}\right)^2\!\left(\rho_{,x}{}^2+\rho_{,y}{}^2\right)\!+\!2R^2(\epsilon+2E)\rho_{,z}{}^2\right]$. The normalization factor $N$ (obtained from the requirement $k_{\mu}l^{\mu}=-2$) reads
\begin{equation}
\label{eq:ss-reparam-N}
N = \frac{R^2\left(R_{,z}-R\frac{\calE_{,z}}{\calE}\right)^2}{\left(R_{,z}\calE-R\calE_{,z}\right)^2\left(\rho_{,x}{}^2+\rho_{,y}{}^2\right)+R^2(\epsilon+2E)\rho_{,z}{}^2} \,,
\end{equation}
and is always positive. The Laplace operator is negative definite, therefore the equation $\theta^{(l)}=0$ has a non-trivial solution if and only if $\calZ<0$. Choosing $R_{,t}<0$, i.e. a collapsing phase of the universe, this means that a horizon exists if $\left(R_{,z}-R\frac{\calE_{,z}}{\calE}\right)_{\!\!,t}>0$. Then $\theta^{(k)}<0$, and so the horizon is future. On this ''side'' of a shell-crossing singularity with $R_{,z}-R\frac{\calE_{,z}}{\calE}<0$, the condition $\left(R_{,z}-R\frac{\calE_{,z}}{\calE}\right)_{\!\!,t}>0$ can be interpreted as the neighbouring matter shells moving towards each other, see Figure \ref{fig:shellcross}. This corresponds perfectly with the universe being in a collapsing phase. On the other hand, the equation $\theta^{(k)}=0$ has a non-trivial solution if and only if $\calZ>0$. Assuming $R_{,t}>0$, i.e. the universe expands, a horizon exists if $\left(R_{,z}-R\frac{\calE_{,z}}{\calE}\right)_{\!\!,t}<0$, and it is past because $\theta^{(l)}>0$. Similarly, the condition $\left(R_{,z}-R\frac{\calE_{,z}}{\calE}\right)_{\!\!,t}<0$ can be interpreted as the neighbouring matter shells moving away from each other, which agrees with an expanding phase of the universe. Moreover, a kind of degenerate horizon given by the conditions $\theta^{(k)}=0$, $\theta^{(l)}=0$, which in the Schwarzschild spacetime corresponds to the Einstein--Rosen bridge, cannot exist in this spacetime.

Let us now consider the case $R_{,z}-R\frac{\calE_{,z}}{\calE}>0$. The future-directed null normals and the corresponding vectors change only slightly compared to the previous case (the only difference is in the time component). One of the null normals reads
\begin{eqnarray}
\label{eq:ss-reparam-kkvect2}
k_{\mu}\ud x^{\mu} =& -\frac{\sqrt{\left(R_{,z}\calE-R\calE_{,z}\right)^2\left(\rho_{,x}{}^2+\rho_{,y}{}^2\right)+R^2(\epsilon+2E)\rho_{,z}{}^2}}
{R\left(R_{,z}-R\frac{\calE_{,z}}{\calE}\right)}\,\ud t \\ 
&+ \rho_{,x}\,\ud x + \rho_{,y}\,\ud y + \rho_{,z}\,\ud z \,, \nonumber \\[0.5cm]
k^{\mu}\frac{\partial}{\partial x^{\mu}} =& \frac{\sqrt{\left(R_{,z}\calE-R\calE_{,z}\right)^2\left(\rho_{,x}{}^2+\rho_{,y}{}^2\right)+R^2(\epsilon+2E)\rho_{,z}{}^2}}
{R\left(R_{,z}-R\frac{\calE_{,z}}{\calE}\right)}\,\frac{\partial}{\partial t} \\
&+ \frac{\calE^2}{R^2}\left(\rho_{,x}\,\frac{\partial}{\partial x} + \rho_{,y}\,\frac{\partial}{\partial y}\right) + \frac{\epsilon+2E}{\left(R_{,z}-R\frac{\calE_{,z}}{\calE}\right)^2}\,\rho_{,z}
\,\frac{\partial}{\partial z} \,, \nonumber
\end{eqnarray}
and the other one is
\begin{eqnarray}
\label{eq:ss-reparam-llvect2}
l_{\mu}\ud x^{\mu} = -\frac{R\left(R_{,z}-R\frac{\calE_{,z}}{\calE}\right)}{\sqrt{\left(R_{,z}\calE-R\calE_{,z}\right)^2\left(\rho_{,x}{}^2+\rho_{,y}{}^2\right)+R^2(\epsilon+2E)\rho_{,z}{}^2}}\,\ud t &\\[0.1cm] 
-\frac{R^2\left(R_{,z}-R\frac{\calE_{,z}}{\calE}\right)^2}{\left(R_{,z}\calE-R\calE_{,z}\right)^2\left(\rho_{,x}{}^2+\rho_{,y}{}^2\right)+R^2(\epsilon+2E)\rho_{,z}{}^2} \left(\rho_{,x}\,\ud x + \rho_{,y}\,\ud y + \rho_{,z}\,\ud z\right)\,, &\nonumber\\[0.7cm]
l^{\mu}\frac{\partial}{\partial x^{\mu}} = \frac{R\left(R_{,z}-R\frac{\calE_{,z}}{\calE}\right)}{\sqrt{\left(R_{,z}\calE-R\calE_{,z}\right)^2\left(\rho_{,x}{}^2+\rho_{,y}{}^2\right)+R^2(\epsilon+2E)\rho_{,z}{}^2}} \,\frac{\partial}{\partial t}& \\[0.1cm]
-\frac{\left(R_{,z}\calE-R\calE_{,z}\right)^2}{\left(R_{,z}\calE-R\calE_{,z}\right)^2\left(\rho_{,x}{}^2+\rho_{,y}{}^2\right)+R^2(\epsilon+2E)\rho_{,z}{}^2}
\left(\rho_{,x}\,\frac{\partial}{\partial x} + \rho_{,y}\,\frac{\partial}{\partial y}\right) & \nonumber\\
-\frac{R^2(\epsilon+2E)}{\left(R_{,z}\calE-R\calE_{,z}\right)^2\left(\rho_{,x}{}^2+\rho_{,y}{}^2\right)+R^2(\epsilon+2E)\rho_{,z}{}^2} \,\rho_{,z}\,\frac{\partial}{\partial z}\,. &\nonumber
\end{eqnarray}
Expansions of the null normals are now
\begin{equation}
\label{eq:ss-reparam-expanzek2}
\theta^{(k)} = \Delta\rho(t_0,x,y,z)-\calZ(t_0,x,y,z) \,,
\end{equation}
\begin{equation}
\label{eq:ss-reparam-expanzel2}
\theta^{(l)} = N\left\{-\Delta\rho(t_0,x,y,z)-\calZ(t_0,x,y,z)\right\} \,,
\end{equation}
with $\calZ$ and $N$ remaining in the same form. On this side of a shell-crossing singularity, a future horizon on which $\theta^{(l)}=0$, $\theta^{(k)}<0$ is satisfied, forms in a collapsing phase of the universe $R_{,t}<0$ if $\left(R_{,z}-R\frac{\calE_{,z}}{\calE}\right)_{\!\!,t}<0$. The interpretation in this case is that the neighbouring matter shells are moving towards each other. On the contrary, a past horizon satisfying $\theta^{(k)}=0$, $\theta^{(l)}>0$, exists in an expanding phase of the universe $R_{,t}>0$ if $\left(R_{,z}-R\frac{\calE_{,z}}{\calE}\right)_{\!\!,t}>0$. This means that the neighbouring shells of matter are moving away from each other. As on the other side of a shell-crossing singularity, a horizon given by $\theta^{(k)}=0$, $\theta^{(l)}=0$ cannot exist on this side.

\subsection*{Special cases}
In this section, we abandon for a moment our efforts to analyse the most general horizon, and simplify things by restricting ourselves using certain assumptions. Firstly, we consider the areal radius $R(t,z)$ to be independent of the coordinate $z$, and secondly, we consider a horizon that respects a quasisymmetry.

\subsubsection*{Areal radius independent of $z$}
Suppose that the areal radius depends only on the time coordinate. Then, to avoid a shell-crossing singularity, we need to consider either $(\ln\calE)_{,z}>0$ or $(\ln\calE)_{,z}<0$. In the first case, expansions of the null normals remain in the form \eref{eq:ss-reparam-expanzek}, \eref{eq:ss-reparam-expanzel}, and in the second case, the expansions keep the form as in \eref{eq:ss-reparam-expanzek2}, \eref{eq:ss-reparam-expanzel2}, while functions $\calZ$ and $N$ now read
\begin{eqnarray}
\label{eq:ss-reparam-spec-ZN}
\calZ & = 2R_{,t}\,\frac{\calE}{R^2\calE_{,z}} \sqrt{\calE_{,z}{}^2(\rho_{,x}{}^2+\rho_{,y}{}^2)+(\epsilon+2E)\rho_{,z}{}^2} \,,\\
N & = \frac{R^2\calE_{,z}{}^2}{\calE^2\left[\calE_{,z}{}^2(\rho_{,x}{}^2+\rho_{,y}{}^2)+(\epsilon+2E)\rho_{,z}{}^2\right]} \,.
\end{eqnarray}
Therefore, by the same argumentation as before, on both sides of the shell-crossing singularity a future horizon $\theta^{(l)}=0$, $\theta^{(k)}<0$ exists only in a collapsing phase of the universe $R_{,t}<0$, whereas a past horizon $\theta^{(k)}=0$, $\theta^{(l)}>0$ forms only when the universe expands, i.e. $R_{,t}>0$. 

\subsubsection*{Quasisymmetric horizon}
We will now assume that the horizon respects a quasisymmetry as in \cite{hellkras,krasbol,kras}, i.e. it is a surface $\{t=\textrm{const},z=\textrm{const}\}$. 

Let us derive the equation of the horizon in the usual way. The equation describing a spatial slice of the horizon now takes the form
\begin{equation}
\label{eq:ss-sym-ro}
\rho(t_0,z) = \mathrm{const} \,.
\end{equation}

As in the general case, we discuss the choice $R_{,z}-R\frac{\calE_{,z}}{\calE}<0$ first. One of the null normal 1-forms and its corresponding vector are then
\begin{eqnarray}
\label{eq:ss-sym-kkvect}
k_{\mu}\ud x^{\mu} & = \rho_{,z}\left(-\frac{\sqrt{\epsilon+2E}}{R\frac{\calE_{,z}}{\calE}-R_{,z}}\,\ud t + \ud z\right) \,,\\[0.1cm]
k^{\mu}\frac{\partial}{\partial x^{\mu}} & = \rho_{,z}\,\frac{\sqrt{\epsilon+2E}}{R\frac{\calE_{,z}}{\calE}-R_{,z}}\left(\frac{\partial}{\partial t} + \frac{\sqrt{\epsilon+2E}}{R\frac{\calE_{,z}}{\calE}-R_{,z}}\,\frac{\partial}{\partial z}\right) \,.
\end{eqnarray}
This is a future-directed vector if and only if $\rho_{,z}>0$. The other future-directed null normal 1-form and the corresponding vector are
\begin{eqnarray}
\label{eq:ss-sym-llvect}
l_{\mu}\ud x^{\mu} & = \frac{1}{\rho_{,z}}\frac{R\frac{\calE_{,z}}{\calE}-R_{,z}}{\sqrt{\epsilon+2E}}\left(-\,\ud t - \frac{R\frac{\calE_{,z}}{\calE}-R_{,z}}{\sqrt{\epsilon+2E}}\,\ud z\right) \,, \\[0.1cm]
l^{\mu}\frac{\partial}{\partial x^{\mu}} & = \frac{1}{\rho_{,z}}\left(\frac{R\frac{\calE_{,z}}{\calE}-R_{,z}}{\sqrt{\epsilon+2E}}\,\frac{\partial}{\partial t} - \frac{\partial}{\partial z}\right) \,.
\end{eqnarray}
The induced metric takes the form
\begin{equation}
\label{eq:ss-sym-h}
h_{\mu\nu}\ud x^{\mu}\ud x^{\nu} = \frac{R^2}{\calE^2}\,\ud x^2 + \frac{R^2}{\calE^2}\,\ud y^2 \,.
\end{equation}
And finally, expansions of the null normals read
\begin{eqnarray}
\label{eq:ss-sym-expanze}
\theta^{(k)} & = 2\rho_{,z}\,
\frac{\epsilon+2E}
{R\left(R\frac{\calE_{,z}}{\calE}-R_{,z}\right)}\left(\frac{R_{,t}}{\sqrt{\epsilon+2E}}-1\right) \,,\\
\theta^{(l)} & = \frac{2}{\rho_{,z}}
\frac{R\frac{\calE_{,z}}{\calE}-R_{,z}}
{R}\left(\frac{R_{,t}}{\sqrt{\epsilon+2E}}+1\right) \,.
\end{eqnarray}
Since all the functions $\rho_{,z}$, $\epsilon+2E$, $R(t,z)$ and $R\frac{\calE_{,z}}{\calE}-R_{,z}$ are positive, the sign of $\theta^{(k)}$ is given by the sign of the expression $\frac{R_{,t}}{\sqrt{\epsilon+2E}}-1$, and the sign of $\theta^{(l)}$ is given by the sign of $\frac{R_{,t}}{\sqrt{\epsilon+2E}}+1$. In a collapsing phase of the universe when $R_{,t}<0$, a future horizon satisfying $\theta^{(l)}=0$, $\theta^{(k)}<0$ exists, and is given by the equation $R_{,t}=-\sqrt{\epsilon+2E}$. On the other hand, when the universe is in an expanding phase, i.e. $R_{,t}>0$, a past horizon satisfying $\theta^{(k)}=0$, $\theta^{(l)}>0$ exists, and is given by the equation $R_{,t}=\sqrt{\epsilon+2E}$. As in the general case, a horizon on which $\theta^{(k)}=0$, $\theta^{(l)}=0$ does not exist.

Let us now analyse the other case when $R_{,z}-R\frac{\calE_{,z}}{\calE}>0$. Again, the future-directed null normals change only in their time component. One of them takes the form
\begin{eqnarray}
\label{eq:ss-sym-kkvect2}
k_{\mu}\ud x^{\mu} & = \rho_{,z}\left(-\frac{\sqrt{\epsilon+2E}}{R_{,z}-R\frac{\calE_{,z}}{\calE}}\,\ud t + \ud z\right) \,,\\[0.1cm]
k^{\mu}\frac{\partial}{\partial x^{\mu}} & = \rho_{,z}\,\frac{\sqrt{\epsilon+2E}}{R_{,z}-R\frac{\calE_{,z}}{\calE}}\left(\frac{\partial}{\partial t} + \frac{\sqrt{\epsilon+2E}}{R_{,z}-R\frac{\calE_{,z}}{\calE}}\,\frac{\partial}{\partial z}\right) \,,
\end{eqnarray}
and the other one reads
\begin{eqnarray}
\label{eq:ss-sym-llvect2}
l_{\mu}\ud x^{\mu} & = \frac{1}{\rho_{,z}}\frac{R_{,z}-R\frac{\calE_{,z}}{\calE}}{\sqrt{\epsilon+2E}}\left(-\,\ud t - \frac{R_{,z}-R\frac{\calE_{,z}}{\calE}}{\sqrt{\epsilon+2E}}\,\ud z\right) \,, \\[0.1cm]
l^{\mu}\frac{\partial}{\partial x^{\mu}} & = \frac{1}{\rho_{,z}}\left(\frac{R_{,z}-R\frac{\calE_{,z}}{\calE}}{\sqrt{\epsilon+2E}}\,\frac{\partial}{\partial t} - \frac{\partial}{\partial z}\right) \,.
\end{eqnarray}
As before, their future orientation is ensured by $\rho_{,z}>0$. The induced metric remains in the same form, and the expansions of the null normals now read
\begin{eqnarray}
\label{eq:ss-sym-expanze2}
\theta^{(k)} & = 2\rho_{,z}\,
\frac{\epsilon+2E}
{R\left(R_{,z}-R\frac{\calE_{,z}}{\calE}\right)}\left(\frac{R_{,t}}{\sqrt{\epsilon+2E}}+1\right) \,,\\
\theta^{(l)} & = \frac{2}{\rho_{,z}}
\frac{R_{,z}-R\frac{\calE_{,z}}{\calE}}
{R}\left(\frac{R_{,t}}{\sqrt{\epsilon+2E}}-1\right) \,.
\end{eqnarray}
In this case, the sign of $\theta^{(k)}$ is given by the sign of $\frac{R_{,t}}{\sqrt{\epsilon+2E}}+1$, and the sign of $\theta^{(l)}$ is given by the sign of $\frac{R_{,t}}{\sqrt{\epsilon+2E}}-1$. Therefore, in a collapsing phase of the universe a future horizon satisfying $\theta^{(k)}=0$, $\theta^{(l)}<0$ exists, and is defined by the equation $R_{,t}=-\sqrt{\epsilon+2E}$. And when the universe expands, a past horizon satisfying $\theta^{(l)}=0$, $\theta^{(k)}>0$ forms, and is given by the equation $R_{,t}=\sqrt{\epsilon+2E}$. Again, a horizon given by $\theta^{(k)}=0$, $\theta^{(l)}=0$ cannot exist.

Finally, let us determine whether the horizons are outer or inner. We will proceed in the same way as when analysing the Lema\^{i}tre spacetime, i.e. we will use the cross-focusing equation. If a horizon (future or past) satisfies $\theta^{(k)}=0$, we use the equation in the form \eref{eq:crossraychaud}, and if a horizon is given by $\theta^{(l)}=0$, we interchange the vectors $k^{\mu}\longleftrightarrow l^{\mu}$. The right-hand side of the equation remains in the form \eref{eq:lem-Tkl}, and the Ricci scalar of the horizons reads $\mathcal{R}=\frac{2\calE^2}{R^2}\left[(\ln\calE)_{,xx}+(\ln\calE)_{,yy}\right]$. Then the horizons are
\begin{equation}
\label{eq:ss-sym-out}
\textrm{outer}\,\quad\Leftrightarrow\,\quad \kappa\varepsilon < \frac{\calE^2}{2R^2}\left[(\ln\calE)_{,xx}+(\ln\calE)_{,yy}\right]+\kappa p  = \frac{2\epsilon}{R^2} + \kappa p\,,
\end{equation}
\begin{equation}
\label{eq:ss-sym-in}
\textrm{inner}\,\quad\Leftrightarrow\,\quad \kappa\varepsilon > \frac{\calE^2}{2R^2}\left[(\ln\calE)_{,xx}+(\ln\calE)_{,yy}\right]+\kappa p = \frac{2\epsilon}{R^2} + \kappa p\,,
\end{equation}
where the sign of $\epsilon$ determines if the horizon section $\{t=\textrm{const},z=\textrm{const}\}$ has spherical, planar or hyperboloidal geometry \cite{plebkras}. Naturally, our definition expects compact horizon sections only.

\begin{table}[h!]
	\centering
	\ra{1.5}
	\begin{tabular}{@{}lclclcl@{}} 
		\toprule[1.5pt]
		\multicolumn{7}{c}{General horizon: $\rho(t_0,x,y,z)=\mathrm{const}$} \\
		\multicolumn{7}{c}{$R=R(t,z)$} \\
		\midrule
		\multicolumn{3}{c}{For a shell-crossing side} & \phantom{a} & \multicolumn{3}{c}{For a shell-crossing side} \\
		\multicolumn{3}{c}{$R_{,z}-R\frac{\calE_{,z}}{\calE}<0$} & \phantom{a} & \multicolumn{3}{c}{$R_{,z}-R\frac{\calE_{,z}}{\calE}>0$} \\
		Future horizon & \phantom{} & Past horizon & \phantom{a} & Future horizon & \phantom{} & Past horizon \\
		\cmidrule{1-1}\cmidrule{3-3}\cmidrule{5-5}\cmidrule{7-7}
		exists when & \phantom{} & exists when & \phantom{a} & exists when & \phantom{} & exists when \\
		$R_{,t}<0$ & \phantom{} & $R_{,t}>0$ & \phantom{a} & $R_{,t}<0$ & \phantom{a} & $R_{,t}>0$ \\
		(Collapsing) & \phantom{} & (Expanding) & \phantom{a} & (Collapsing) & \phantom{} & (Expanding) \\[0.1cm]
		$\left(R_{,z}-R\frac{\calE_{,z}}{\calE}\right)_{\!\!,t}\!>\!0$ & \phantom{} & $\left(R_{,z}-R\frac{\calE_{,z}}{\calE}\right)_{\!\!,t}\!<\!0$ & \phantom{a} & $\left(R_{,z}-R\frac{\calE_{,z}}{\calE}\right)_{\!\!,t}\!<\!0$ & \phantom{} & $\left(R_{,z}-R\frac{\calE_{,z}}{\calE}\right)_{\!\!,t}\!>\!0$ \\[0.2cm]
		\stackanchor{(Matter shells}{approaching)} & \phantom{} & \stackanchor{(Matter shells}{moving away)} & \phantom{a} & \stackanchor{(Matter shells}{approaching)} & \phantom{} & \stackanchor{(Matter shells}{moving away)} \\
		\bottomrule[1.5pt]
	\end{tabular}
	\caption{Horizon existence in the Szekeres--Szafron spacetime -- general case}
	\label{tab:ss-general}
\end{table}

\begin{table}[h!]
	\centering
	\ra{1.5}
	\begin{tabular}{@{}lclclcl@{}} 
		\toprule[1.5pt]
		\multicolumn{3}{c}{General horizon: $\rho(t_0,x,y,z)=\mathrm{const}$} & \phantom{a} & \multicolumn{3}{c}{Quasisymmetric horizon: $\rho(t_0,z)=\mathrm{const}$} \\
		\multicolumn{3}{c}{$R=R(t)$} & \phantom{a} & \multicolumn{3}{c}{$R=R(t,z)$} \\
		\midrule
		\multicolumn{3}{c}{For both shell-crossing sides} & \phantom{a} & \multicolumn{3}{c}{For both shell-crossing sides} \\
		\multicolumn{3}{c}{$(\ln\calE)_{,z}\lessgtr0$} & \phantom{a} & \multicolumn{3}{c}{$R_{,z}-R\frac{\calE_{,z}}{\calE}\lessgtr0$} \\
		Future horizon & \phantom{} & Past horizon & \phantom{a} & Future horizon & \phantom{} & Past horizon \\
		\cmidrule{1-1}\cmidrule{3-3}\cmidrule{5-5}\cmidrule{7-7}
		exists when & \phantom{} & exists when & \phantom{a} & exists when & \phantom{} & exists when \\
		$R_{,t}<0$ & \phantom{} & $R_{,t}>0$ & \phantom{a} & $R_{,t}<0$ & \phantom{} & $R_{,t}>0$ \\
		(Collapsing) & \phantom{} & (Expanding) & \phantom{a} & (Collapsing) & \phantom{} & (Expanding) \\
		\bottomrule[1.5pt]
	\end{tabular}
	\caption{Horizon existence in the Szekeres--Szafron spacetime -- special cases}
	\label{tab:ss-special}
\end{table}

\section*{Conclusions}
We have studied quasilocal horizons in two inhomogeneous cosmological models.

In the spherically symmetric Lema\^{i}tre spacetime we discovered a future and a past horizon given by equations \eref{eq:lem-rcehor}--\eref{eq:lem-minhor}. We confirmed our expectations that the future horizon exists only in a collapsing phase of the universe, whereas the past one exists only when the universe expands. Using adapted coordinates, we found out that the matter on the horizons behaves as a perfect fluid with negative pressure provided that the Misner--Sharp mass is constant along the horizons. Under the same assumption we proved that both horizons are null and non-expanding according to the definition by Ashtekar and Krishnan \cite{ashtekar}. We also found the conditions under which they are outer or inner \eref{eq:lem-out}--\eref{eq:lem-in}. Finally, comparing intrinsic and extrinsic horizon geometry in the Lema\^{i}tre and Lema\^{i}tre--Tolman--Bondi spacetime, we came to the conclusion that both spacetimes have locally the same geometry of the horizons.

We have also studied a generalization of the Lema\^{i}tre spacetime, namely a subfamily of the Szekeres--Szafron spacetime with the metric function derivative $\beta_{,z}$ non-vanishing. This spacetime has no symmetries, therefore we first assumed that the horizon did not have any symmetries either. We were able to find certain conditions on the horizon existence, namely that a future horizon exists in a collapsing phase of the universe when the neighbouring matter shells are moving towards each other. On the other hand, a past horizon exists in an expanding phase of the universe when the neighbouring matter shells are moving away from each other. We have also analysed two special cases: the areal radius independent of the coordinate $z$ and the horizons respecting a quasisymmetry. We found similar conditions on the horizon existence as in the general case apart from the requirements for the neighbouring matter shells distance. Moreover, in the case of a quasisymmetric horizon we determined when the horizons are outer and inner \eref{eq:ss-sym-out}--\eref{eq:ss-sym-in}. All the results in the Szekeres--Szafron spacetime are summarized in Table \ref{tab:ss-general} and \ref{tab:ss-special}.

\section*{Acknowledgements}
The research was supported by the Charles University, project GAUK No. 369015 (E.P.) and by GAČR No. 17-13525S (E.P. and O.S.).

\end{document}